\documentclass[graybox]{svmult}

\usepackage{type1cm}        
\usepackage{makeidx}         
\usepackage{graphicx}        
\usepackage{multicol}        
\usepackage[bottom]{footmisc}

\usepackage{newtxtext}       %
\usepackage{newtxmath}       

\makeindex             

\begin{document}

\title*{Teaching and Learning Ethnography for Software Engineering Contexts}
\author{Yvonne Dittrich, Helen Sharp and Cleidson de Souza
\\
A slightly revised version of the text is published in: 
\\
Daniel Méndez, Paris Avgeriou, Marcos Kalinowski, Nauman Bin Ali:
Handbook on Teaching Empirical Software Engineering. Springer Nature Switzerland 2024, ISBN 978-3-031-71768-0}
\authorrunning{Yvonne Dittrich, Helen Sharp, Cleidson de Souza}
\institute{Yvonne  Dittrich \at IT University of Copenhagen, Rued Langgaards Vej 7, 2300 Copenhagen S, Denmark \email{ydi@itu.dk}
\and Helen Sharp \at The Open University, Walton Hall, Milton Keynes, MK7 6AA, UK \email{helen.sharp@open.ac.uk}
\and Cleidson R. B. de Souza \at Universidade Federal do Pará: Belém, Pará, BR \email{cleidson.desouza@acm.org }
}
%
%
\maketitle

\abstract*{
Ethnography has become one of the established methods for empirical research on software engineering. Although there is a wide variety of introductory books available, there has been no material targeting software engineering students particularly, until now.\\
In this chapter we provide an introduction to teaching and learning ethnography for faculty teaching ethnography to software engineering graduate students and for the students themselves of such courses.
\\The contents of the chapter focuses on what we think is the core basic knowledge for newbies to ethnography as a research method. We complement the text with proposals for exercises,  tips for teaching, and pitfalls that we and our students have experienced.
\\The chapter is designed to support part of a course on empirical software engineering and provides pointers and literature for further reading. }

\abstract{
Ethnography has become one of the established methods for empirical research on software engineering. Although there is a wide variety of introductory books available, there has been no material targeting software engineering students particularly, until now.\\
In this chapter we provide an introduction to teaching and learning ethnography for faculty teaching ethnography to software engineering graduate students and for the students themselves of such courses.
\\The contents of the chapter focuses on what we think is the core basic knowledge for newbies to ethnography as a research method. We complement the text with proposals for exercises,  tips for teaching, and pitfalls that we and our students have experienced.
\\The chapter is designed to support part of a course on empirical software engineering and provides pointers and literature for further reading. 
}

\section{Introduction}
\label{sec:1 Introduction}

Ethnographic studies are an important element of the toolbox for empirical software engineering as they can provide unique insights into software development practice \cite{sharp2016role}.
Ethnography was originally developed as a method to understand foreign cultures by becoming embedded in them and documenting the learning and sense-making that was experienced during the process \cite{hammersley2019ethnography}. Modern ethnography has been developed in an attempt to understand a foreign culture from a members’ perspective. With a growing interest in the understanding of subcultures close to the ethnographer, ethnographic research was used to learn more about a wide range of contexts including truck drivers \cite{agar1986independents}, midwives \cite{jordan1989cosmopolitical}, London underground workers \cite{DBLP:conf/ecscw/HeathL91} and organisational life \cite{ DBLP:journals/tem/StrodeSBGT22,Ybemaetal2009}.

In the context of Human-Computer Interaction and Computer Supported Cooperative Work, ethnographic research was used to understand how the members of an organisation made use of technology. Examples for this kind of research include several studies of air traffic controllers in the context of a redesign of the system supporting their perception and control of air traffic \cite{bentley1992ethnographically}, of copy machine repair personnel by Julian Orr \cite{orr1990talking}, and of attorneys and paralegals to prepare the design of software supporting their document management \cite{blomberg1996reflections}.
For more than 25 years, ethnography has also been used in the context of software engineering research in order to study cooperative aspects of software engineering. Examples are a study on configuration management \cite{DBLP:conf/icse/HerbslebG99}, a study on the use of social software in distributed software development \cite{GiuffridaPhD, DBLP:conf/icse/GiuffridaD14}, and an ethnographic study of a small software house to investigate testing practices \cite{DBLP:conf/icse/MartinRRS07}. 
As in the original context, the benefit of ethnographic studies of software engineering is the understanding of how and why software teams do what they do to develop software. Ethnography allows understanding and describing collaboration in software engineering from the software engineers’ perspective. Ethnographic studies require the researcher to join a development team with the intent to understand the why and how of the team’s everyday practices. 

\begin{tips}{The Educators’ Corner: Tips for teaching}

\begin{itemize}
    \item It’s important for students to know the origins of ethnography but introduce the approach through a mixture of examples from software engineering and other disciplines so that the relevance to software engineering comes through
    \item If you don’t have (much) experience of ethnography yourself then identify colleagues with experience to engage in your classes
    \item It’s hard to see another person’s point of view, offer your students guidance on emotional intelligence, and how to see different perspectives
    \item Present examples of existing ethnographies in software engineering for the students to facilitate their writing
    \item Take an iterative approach to topics, e.g. provide some background, set an exercise, discuss the exercise and ask students to do the exercise again
    \item Embrace the use of ethnography with other techniques, e.g. interviews, as these are often required in the field
    \item Because activity in the field is fluid, ethical dimensions of the study need careful attention 
\end{itemize}

\end{tips}

\noindent When we use the term ‘practices’ in this chapter, we talk about the established way a software team goes about developing and evolving software. Social practices, like cooperative software engineering, are based on an understanding of relevant objects, representations and tasks, on implicit and explicit rules, and on a common goal (for further readings see Section \ref{sec:7}). 

Understanding software practices using an ethnographic approach aims to understand how and why the software development practices make sense from a members’ point of view, putting aside ideas on how to improve the observed practices for the time being. This can be hard for a software engineering researcher. As engineers and software engineering researchers we are trained to design and create things and to improve the methods used to this end. We tend to focus on problems and short-comings, rather than aiming to understand how a complex activity such as software development actually takes place, or what enables a team to produce software at all. Ethnographic research invites us to temporarily put aside or ‘bracket’ \cite{Gearing} our assumptions of how software development should take place, and try to understand how developers make things work no matter what. 

In return, the results of ethnographic research help us understand the rationale behind even seemingly disadvantageous behaviour, which in turn provides us with important information when designing tools and methods. An example can be found in \cite{Unphon2009, UphonDittrich2008}. The article presents an interview study triangulating an ethnographic study of software architecture practices for a software product. The researchers were surprised at the lack of any software architecture documentation and the unwillingness of the team to produce and maintain such documentation. A triangulating interview study revealed that this reluctance towards software architecture was shared by architects and tech leads of other products. The only exception, an open source project, also provided us with a rationale for this attitude: the existence of architecture documentation can prohibit discussion between the architects and the developers with the effect that the architects might not know how the developers change the architecture or where the architecture creates problems for the development and might need to be revised. This, in turn, means that tools for designing and communicating software architecture should be designed to support architect-developer discussions and not replace them. Had we not taken the reservations of the developers and architects seriously, we would not have understood that there was a ‘good reason for bad architecture documentation’ \cite{UphonDittrich2008}.

Ethnographic studies can also lay the groundwork for practitioner support tools that are accepted by the community because they are based on an understanding of practitioners' point of view. An example of this is the motivating jenny project which ran a multi-site ethnographic study to understand developers’ security behaviour. This project resulted in a number of insights including a model of security behaviour \cite{DBLP:journals/tosem/LopezSBTLN23} that underpins a set of four practitioner packs designed to improve the security culture in a team or organisation (downloadable from motivatingjenny.org). These have been downloaded thousands of times and have been used in education and in industrial settings. 

Our previous, more theoretical and programmatic article in Transactions on Software Engineering \cite{sharp2016role} provides examples from empirical software engineering where ethnography has been used. This chapter complements that article with support on how to implement and teach ethnographic research in a Software Engineering context. The goal of this chapter is twofold. On the one hand, we introduce ethnography as a research method for software engineering with a hands-on approach. We aim to provide concrete how-to support for researchers starting to use ethnography. Our intended reader in connection to this goal is a researcher, e.g. a PhD student, who wants to explore ethnography as a possible research method. On the other hand, we provide guidance for teachers of a postgraduate course on empirical research in software engineering. The chapter comprises what we see as the minimum contents of a module on ethnography as an empirical method for software engineering. 

In our tutorials, people have often asked us what is the difference between ethnography and other qualitative methods, such as contextual inquiry and grounded theory. And why not just do interviews? A simple response to these questions might include that contextual inquiry is based on an apprenticeship model, grounded theory seeks concepts in the domain that can be developed into a theory, and in interviews participants’ accounts will necessarily be partial and rationalised. Ethnographic research, on the other hand, focuses on a holistic view of the practice being studied, has research questions that evolve over the course of the study, and aims to capture activity as it happens together with practitioners' rationale. But there is more to it than this. This chapter will provide a deeper answer to these questions.

The remainder of the chapter is structured as follows: Section 2 discusses whether ethnography is the right choice from three perspectives: the role of ethnographic studies in relation to your specific research context; the type of research question it can address; and what is required of the researcher. Section 3 discusses the planning of an ethnographic study. Section 4 covers important topics such as field work and addresses issues ranging from the role the researchers take on in relation to the team, to how to design and keep a field diary. Section 5 introduces the analysis of ethnographic data and section 6 covers research ethics for ethnography. Section 7 concludes the chapter and provides suggestions for further reading. 

\begin{question} {Exercise 1: Identify an ethnographic research project}

\noindent The exercises in this chapter together result in the design of an ethnographic study project. If you already plan to do ethnographic research, feel free to use your project throughout the exercises. 

If you use other methods in your research, try to find an angle to your research that could benefit from a real world investigation. For example, if you work with mining software repositories (MSR), you might consider conducting an ethnographic study of one of the projects that would shed light on the data you base your MSR study on.

Formulate a short text (about half a page) motivating your study and formulating a research question related to understanding relevant development practices from a members’ point of view.

\end{question}
\begin{tips}{The Educators’ Corner: Exercises}

\noindent Throughout the chapter, we propose a number of hands-on exercises that encourage students to apply the contents in a concrete situation. Whereas the introduction to ethnography targets both students and more senior researchers, the exercises are directed to a student designing their empirical research project. The order of the sections in this chapter is designed to support a logical progression in these exercises. 

The exercise for Section 2 asks students to formulate an adequate research question for their own or a fictional ethnographic study. Section 3 discusses the planning of the ethnographic study, accompanied by an exercise to develop a preliminary plan for their own or a fictional study. Section 4 describes the field work. Here we propose an exercise that sends the students out to observe an everyday situation to gain hands-on experience. In a second exercise, the students are asked to prepare the concrete field work. Section 5 discusses the analysis of the field material. An experienced researcher might bring their own empirical data to the course and provide the students with hands-on experience with field material. As a fall back exercise we propose the development of an analysis strategy for their own or a fictional research project. The discussion of ethnographic research ethics in section 6 should enable the students to review their study designs from an ethics perspective and discuss the ethical challenges of their research design. Section 7 does not have specific exercises. 

\end{tips}

\section{Is an ethnographic study the right choice?}
\label{sec:2Choice}

Here we consider the question of whether or not an ethnographic study is the right choice from three different perspectives:
\begin{enumerate}
    \item  the context of your research
    \item the kind of research question you want to answer
    \item what it means for you as the researcher, because in an ethnographic study the researcher is the research instrument
\end{enumerate}

\noindent This section will discuss these three perspectives to help you to decide whether to use ethnography as a research method.

\subsection{The context of your research}
This perspective considers where the intended study fits in the wider context: what role might an ethnographic study play in the research? In social sciences, where ethnography originated, studies focus on understanding phenomena. They are used to understand better how (sub) cultures operate, or how groups of people organise their social life. In HCI, ethnographic studies are used to inform the design of technology from a user’s point of view so that the designers might better appreciate who they are designing for. In software engineering, ethnographic studies may also be used to understand or to inform design, but research often goes beyond understanding towards improving methods, techniques and tools. Understanding current practice is a good starting point to inform changes of any kind, not so that current practice can be replicated in the new order, but to appreciate why things are done the way they are, so that there are no unintended consequences when things change. A quite famous example is the long time it has taken to replace the paper flight strips used by air traffic controllers with digital versions. In 1999 Wendy MacKay commented that “Our observations have convinced us that we do not know enough to simply get rid of paper strips, nor can we easily replace the physical interaction between controllers and paper strips” \cite{DBLP:journals/tochi/MacKay99}. She wrote this after being inspired by her work and previous studies about flight strips which started in 1992. A blog written in 2017 on the subject demonstrates how cautiously this change was eventually made \cite{NATS2017}. 

In our analysis of ethnographic research in empirical software engineering \cite{sharp2016role}, we identified four roles for ethnographic research, all of which start by considering “how things work in practice”. 

\subsubsection{\textbf{To strengthen investigations into the social and human aspects of software engineering}}
Since ethnography’s origins are in understanding cultures and communities, it is not a surprise that ethnography has been used in this way. It is inevitable that some insights regarding social and human aspects will emerge whatever research question is pursued since the approach champions the members’ point of view. 

For example, Lopez et al. \cite{DBLP:journals/tosem/LopezSBTLN23} were interested in investigating how non-specialist software developers engage with security in practice. This explicitly focuses on the human aspect of software developers. They used a multi-sited study and a mixture of research methods. Their results included a set of episodes where non-specialists engaged in security activity, and five typical behaviours that characterise how individual developers respond to security concerns to meet the demands of particular circumstances. A key characteristic of this work was that it emphasised security from the developers' perspective. The results form a framework that managers and teams can use to recognize, understand, and alter security activity in their environments (motivatingjenny.org).

\subsubsection{To inform the design of software engineering tools}

For a new tool to be successful it must support the task being performed, e.g. bug localisation. But for it to be useful to practitioners it must also fit into their workflow and support how they and their organisation performs the task. Focusing on a task in isolation gives only a partial view and however clever it may be, if it’s not usable then it won’t be used. So understanding the context of tool use is as important as understanding its technical and functional requirements.

But there are other considerations too. As with much software development, it can be easy to jump straight into designing a system that seems to address a user’s problem \cite{DBLP:books/wi/RSP2023}. However if the goal is to develop new software engineering tools then don’t start the design until you are sure that the proposal will be useful. This can be achieved by sketching a number of potential tools based on observations, and then collecting evidence to support or refute their value, e.g. by working with practitioners and analysing your data overall to evaluate the ideas.  For example deSouza et al. \cite{de2003management}were interested in the role of application programming interfaces (APIs) in the coordination of software development work in large scale organisations. An initial observation led them to think that integrating the versioning software with email might facilitate the team’s work, but further data collection made it clear that this was only part of the situation, and focusing on a tool to support the team’s overall communication would bring them more benefit. So instead of developing something to integrate two existing tools, they developed a tool that supported the identification of dependencies between developers which allowed developers to coordinate their efforts more effectively and reduce unnecessary work.

\subsubsection{To improve development processes}
Ethnography can be a starting point, together with the practitioners observed,  to discuss and implement improvements of their work practices. Ethnographic research then becomes part of action research. Two chapters in this book by Staron and by Dittrich, Bolmsten and Seilein introduce action research. A critical point is that the decision about the intervention and the evaluation of its implementation should keep the members’ point of view, both for ethical and methodological reasons. Dittrich et al. developed Cooperative Method Development (CMD), an action research approach combining ethnographical and ethnomethodological inspired empirical research with the improvement of software engineering tools, methods, and processes \cite{DBLP:journals/ese/DittrichREHL08, Dittrich2002}. CMD is explicitly designed to keep both the work practice focus and the developers’ point of view of the initial research throughout the action research cycle(s). 

Unphon’s thesis provides an example of action research \cite{Unphon2009}. The research cooperation with a company developing software products for simulating hydraulic systems focused on the evolvability of these software products. The researchers collaborated with the team responsible for the product simulating open one-dimensional water systems like rivers and creeks. The action research introduced light-weight software architecture techniques for high-level design, developed a light-weight Architecture Level Evolvability Assessment method for focussed discussions of design decisions with relevant stakeholders, and introduced light-weight architecture compliance techniques using the built system. Introductory research with the team in order to understand development practices as well as investigate the structure of the existing software product resulted in a framework for understanding the influence of the organisation of software development and business on the architecture of the software \cite{UphonDittrich2008}. The research results emphasise the need to adapt software architecture methods and tools to support the continuous evolution of software products: architecture design and evolution take place as part of everyday evolution; architectural practices need to support the software architect to keep up with the changes of the software and the emerging requirements that might challenge the architecture; evolvability should be a quality to be considered in regular software architecture design discussions.

\subsubsection{To inform research programmes that do not have ethnography at their core}
It is common for research outputs such as tools and new processes to be ignored by practitioners, in favour of innovations suggested by other practitioners. Understanding the context of software engineering takes the researcher one step nearer to suggesting an innovation that is acceptable to practice. In particular, an ethnographic study may help a research programme by articulating more specific, relevant research questions. 

For example, our programme on agile software development started by looking at the realities of XP in practice, i.e. how was the approach implemented \cite{DBLP:journals/ese/SharpR04}? Ethnographic studies in this context led to other studies and research questions that focused on the role of physical artefacts \cite{sharp2009role}, information flow between stakeholders \cite{DBLP:journals/ijmms/ZainaSB21} and collaboration in dispersed teams \cite{DBLP:conf/icis/DeshpandeSBG16}.

But ethnography can also be used to inform research programmes by providing context grounded in practice for an existing research focus. For example, Capiluppi et al. \cite{capiluppi2007empirical} investigated the evolution of code within an agile development project. This led to a number of observations regarding how the agile code base evolved. Using an ethnographic study as context allowed the researchers to :
\begin{itemize}
    \item Characterise the development process and practices clearly, and
    \item Provide alternative interpretations of phenomena found in the data, e.g. where the size of the code base changed significantly they could trace an event through the ethnographic study to explain that a new library was loaded
\end{itemize}

Ethnographic observation in this context has the benefit that it is not dependent on the report of practitioners, as it is in interviews. When collecting reports from practitioners, their responses will inevitably be partial and informed by current events. For example, they might not consider certain aspects of everyday practice to be relevant to the questions asked, or they might not report practices that they consider to be informal or not accepted. However, these aspects may be exactly the characteristics that render the specific way of developing software possible, and hence would be important from a research point of view.
The design of studies complementing and providing context to other research may not be driven by an independent research question, but the research question might be decided based on the overall research interest. 

\begin{question}{Exercise 2: What role do you see for your ethnographic study?}

\noindent Consider the four roles above and assess your own context against each one. To which does it fit most closely? It’s ok to align with more than one at this stage. The detailed design will help you narrow it down.

Describe the role that you see for your ethnographic study in a short paragraph and add it to the study design started in Exercise 1.
  
\end{question}
\begin{warning}{The Educator's Corner: Pitfalls}
\begin{itemize}
    \item When conducting an ethnographic study with the aim of investigating social and human aspects of software engineering it is easy to become overwhelmed by the huge number of focus points SO track the possible foci and address them one at a time.
    \item When conducting an ethnographic study with the aim of producing a new tool, it can be tempting to start designing before a full picture has been obtained SO start by hypothesising what kind of tools would be useful and collect data to validate those hypotheses.
    \item When conducting an ethnographic study with the aim of producing a new tool, it can be tempting to focus on specific instances rather than the broader picture SO focus on the analytical results rather than the data when hypothesising about potential tools.
    \item When conducting an ethnographic study with the aim of improving development processes it can be easy to suggest changes based on textbook versions of processes SO make sure to engage practitioners in any suggested modifications.
    \item When conducting an ethnographic study with the aim of informing other programmes of research, confirmation bias can easily cloud your observations SO be sure to seek confirming and disconfirming evidence.
\end{itemize}

\end{warning}

\subsection{The kind of research questions you want to answer}

This subsection considers the kind of research questions you want to answer. The specific detail of the question may change as the study progresses (see section 4 below) but whether an ethnographic study is the right choice depends on the kind of question you want to answer.

The first thing to note is that ethnographic research, like other qualitative research, is driven by research questions rather than hypotheses derived from theory. Research questions ask ‘How’ and ‘Why’ and ‘What are the characteristics of’ questions rather than ‘Is X better than Y’ or ‘Will this technique make programmers more productive?’ kind of questions. For example ‘How do software practitioners develop systems using XP?’ rather than ‘Is single programmer coding more productive than pair programming?’, and ‘Why don’t developers adhere to the company’s security policies?’ rather than ‘Does structuring the manual in this way help developers produce more secure code?’ and ‘What are the characteristics of a technology adoption?’ rather than ‘How did the ideas of Simula develop into Java?

Other techniques and methods introduced in this book will support answering other types of question.

The strength of an ethnographic approach is that it emphasises the point of view of the participants, i.e. the members of the community under study, who are often called informants or interlocutors in ethnographic work. It therefore allows the researcher to understand better why things are the way they are. It brings research closer to practice and hence can make the results more acceptable within practitioner communities. For example, it may seem strange to a researcher to see agile practitioners using both a physical storyboard and a virtual one, especially because this entails duplicate work in keeping them both up to date. However, this is common practice where co-located or hybrid teams are deployed because of the different and complementary collaboration and communication benefits it provides \cite{dingsoyr2023longitudinal}.  

\begin{question}{Exercise 3: What type of research question do you want to answer?}

\noindent Review your research question(s) from Exercise 1. Are they “how” “why” and “what characteristics” questions, as discussed above. If not, can they be recast appropriately?

Revise your study design if needed, and make sure to adjust the whole text to any new formulation of your research question(s).

\end{question}

\subsection{What ethnographic studies require from the researcher}

To help decide whether or not an ethnographic study is the right choice, this section considers the four main features of ethnographic work: the member’s point of view; the ordinary detail of life as it happens, the analytic stance and thick descriptions. For each of these we present considerations on whether the researcher (you) are prepared and willing to apply the approach appropriately, e.g. to write thick descriptions, listen to the informants’ point of view and modify your research focus. The information in this section will help to assess this, and will set your expectations for what it means to implement an ethnographic study. 

\subsubsection{The members’ point of view}

We would argue that it is always important to understand the point of view of practitioners whatever is the focus of the research, for example whether developing a new tool, looking to improve the development process, or simply trying to understand better how something works in practice. As discussed in Section 2.1, taking account of the members’ point of view doesn’t restrict innovation, but instead understanding the rationale for the current way of working before introducing changes helps to avoid unintended consequences. Understanding the member’s point of view is all about understanding what is important for the informants so that this can be taken into account. 

In other disciplines, ethnographic studies involve the ethnographer studying a culture that is unfamiliar to them, e.g. in terms of language, customs, cultural norms etc. These studies may need to last for months or years. As a software engineer conducting an ethnographic study in a software development environment, this will be less of a challenge to overcome. Understanding someone else’s perspective is still hard, but much of the basic context of software development, e.g. IDEs, development processes, programming languages etc will be familiar. One of the benefits of this is that ethnographic studies don’t necessarily need to take months or years. Focusing on the members’ point of view will entail getting used to the organisation’s environment and the specific details of the team’s work but the learning curve is less steep than it would be for someone unfamiliar with software development. 

The familiarity of the researcher with the tools, techniques and methods of the informants can make it difficult to keep the member’s point of view in mind when observing how software development takes place. It is very easy to apply the knowledge from your own education and research and slide into a “this can be done better” attitude. Such an attitude might even be provoked by the practitioners asking the researcher, who supposedly is an expert, whether they have any recommendations. Here it is important to separate observations from analysis for the duration of the field work and put your own judgement aside. 

\subsubsection{The ordinary detail of everyday life}

The ordinary detail of everyday life is important because it exposes how tasks are addressed, the issues that are relevant or not and how informants approach tasks. For example, a series of studies with agile software development teams exposed that the colour of cards placed on a physical scrum board, and the way in which they are handled carries meaning beyond what was written on them \cite{DBLP:journals/ijmms/SharpR08}. This raised questions of whether digital storyboards would support the same kind of processing and collaboration as physical boards.

This also means that the research will be conducted in the field or “in the wild” rather than in a controlled environment, and the research will aim to not disturb or control normal behaviour.

Equally, informants’ language is relevant. For example if the informants’ native language is different to the ethnographer, what might the impact be? If the domain of study is highly technical then the ethnographer would do well to understand the domain too. Performing an ethnographic study in an unfamiliar technical domain will have an impact on the timescale (as mentioned above) but also may lead to misinterpretation of activities and tasks. The gender of the researcher might both influence the field situation and the interpretation of the observation. 

Given that ethnography emphasises what is important from the members’ point of view, and that the researchers won’t know what they might find before entering the field, there is a strong chance that the research focus will evolve. A key issue to consider is whether the research context and design is suitably flexible to allow for the research focus to change? For example, if your focus is developing a new tool to identify areas of technical debt but your informants explain that their concerns are driven by improving the security of the code, how would that affect the research?

Although the detail of everyday life is observed, it is not the detail itself that is important, but the significance of the detail (see section 2.3.3 below).

\subsubsection{The analytic stance}

An ethnographic account is not just a description of what is seen but is crafted out of an analysis and interpretation of what was found in the field, explicating why things are the way they are. The analytic stance is related to the members’ perspective. The idea is to understand and uncover the rationalities of practices and how the practitioners’ behaviour makes sense, even though it does not match the researcher’s own understanding of how software development should take place. This can be problematic to those unused to the approach. 

One example to illustrate this analytical stance can be found in \cite{DBLP:journals/cscw/RonkkoDR05}, where a recording of a steering group meeting addressing a major change in a method and tool development project is analysed. The analysis shows in detail how the steering group relates the actions to take the development further to both the new plan for the project and the company-wide project model. The analysis shows that the steering group takes a decision to deviate from the project model in order to assure that the work in the project can progress using, nonetheless, the very same company-wide project model to make the deviation visible and accountable in the organisation. The analysis thus sheds light on the rationale behind deviations from plans and company-wide project models, and, at the same time, it shows the importance of a company-wide project model to communicate both behaviour according to it and deviations from it to other actors in the organisation.

Not every analysis needs to be as fine grained as the interaction analysis performed in that article. However, the analytical stance requires the researcher to be able to relate the observations here and now to the spatial and temporal context and to the goal and purpose of the observed activities. 

\subsubsection{Thick descriptions}

An ethnographic study results in a comprehensive and detailed set of data, and a detailed account of the findings. The account aims to communicate a broad picture of the study environment and activity. It needs to show how the researcher has arrived at their conclusions or recommendations. The set of data is specific to that site and hence is not suitable for statistical generalisation although it may be suitable for analytical generalisation. If you decide to undertake an ethnographic study this is an important point to remember. In an area where generalisation and statistical significance of results are expected, having a core ethnographic study may be challenging to convince others. The thick description can be used to defend the study and its results, but it cannot justify a statistical generalisation. An ethnographic researcher needs to acquire a different perspective on justification. This may be particularly pertinent for publication and if you are a PhD candidate.

\begin{question}{Exercise 4: Are you ready for this kind of research study?}
\noindent Is there anything in this section about what it means to do an ethnographic study that surprised you, or that seems too difficult? Consider each of these features and write a list of advantages and a list of disadvantages from your point of view. Review these lists and refer to them as you work through the rest of this chapter and the design of your own study.

\end{question}

\section{Planning an ethnographic study}
\label{sec:3Planning}

Ethnography belongs to the category of flexible research designs \cite{RobsonMcCartan}, where both the research focus and the concrete field work are expected to be adapted during the research process. However, this does not imply that planning is redundant. An ethnographic study still needs to be planned. It is important that both the researcher and the practitioners have a shared understanding about what might happen once the researcher joins the project. In addition, the plans should be regularly revisited and adapted, and everybody kept aligned throughout the research.

\subsection{Finding a site for field work}

One of the first things is to identify a suitable project to collaborate with. Often, the ethnographic study would have been prepared by prior contact between the researcher, the company and relevant members of the development team. In other cases, the ethnographic researcher might identify a company or an open source project that fits with the research interest and question. 

When working with a company, the ethnographic study might affect their work, so the company and the project team need to agree to the study and the plan. As ethnography is an extremely flexible way of doing research, good communication with the project and in many cases the management of the company is needed to allow for coordination and replanning. One way of doing this is to create a ‘steering committee’ for the research collaboration, consisting of the researcher, the PI of the project, a project contact, and the person responsible for the research collaboration in the company. This committee would meet regularly to discuss whether changes to the logistics of the field work are needed, and to make sure that resources necessary for the ethnographic study (access to documents, code and the like, time for interviews etc) are provided. Exactly how this group works will depend on the field site circumstances but regular alignment across all parties will keep the study on track.

In other cases, the researcher could decide to become part of an open source community. This in turn will require deciding on the role in the community the researcher would want to take. Are you able to and interested in participating in the Open Source development, i.e. coding? Could you contribute in other ways to the community? Open source members, especially if they take on core roles like maintainers, are often as busy as members of corporate software engineering teams. Either your contribution to the community during the field work or the research results could be received as a return for the time the community members spend with you.

Planning the ethnographic study needs to take research ethics into account as well. Due to the complexity of this topic, research ethics are discussed on their own in Section 6.

The following discusses five dimensions to consider when planning an ethnographic study, especially in software engineering.

\subsection{Participant or non-participant observation }

The best way to understand the culture and practices of a community under study, is to become part of that community. Studying software engineering practices as a software engineer often allows for participation. For example, the researchers can become members of open source communities by participating in the development. However, commercial companies may not be so happy for researchers to be modifying their code, and highly specialised software – for example software for modelling hydraulic systems \cite{Unphon2009} – requires an expertise that software engineering researchers would normally not master. In such cases, the researcher needs to find a role that allows them to observe and become a culturally competent member of the community. One way could be to act as a newly employed project member, an apprentice in the specific development context. In the above case of hydraulic simulation software, that required expertise in differential equations on a PhD level, the field worker took on documenting the architecture of the re-engineered software to support its usage and customisation. Another example was in an early ethnographic study where the researcher took on an administrative role \cite{DBLP:journals/infsof/LowJHHRRW96}. Both of these examples resulted in the researcher getting embedded in the work of the team in a way that was acceptable to all concerned. Deciding the role of the researcher in the field needs to be taken together with the collaborating project team. Also other stakeholders like management and in some cases even clients of the organisation need to be consulted.

It will not always be possible to participate in the development and a more passive observation may be the only option. In this case the researcher will sit quietly and observe, listen and take notes but then it is important to keep alert for particularly pertinent activity. 

\subsection{Duration of field work}

How much time does the researcher need to spend with the software team? In our presentation of published ethnographic studies \cite{sharp2016role}, there was a wide variety of field work durations. The traditional answer of ethnographers would be: until the researcher can act as a competent member of the culture, or until the researcher stops learning new things. As in other qualitative research methods, saturation of the observations is an indication for ending an ethnographic study. Here, the overlap of expertise when researching software engineering practices as a software engineering researcher might shorten the time necessary: although the specific software practices might be new, the field worker shares a common ground with the observed teams in the form of shared professional expertise. However, saturation might not be possible, e.g. if a software development project ends before saturation is reached. On the other hand, iterative processes such as agile software development might allow repeated observation of all relevant practices in a comparatively short time. 

In many situations, it is not necessary to spend all the time with the observed projects. For example, the field worker might choose to spend 3 or 4 days each week with the project, which enables the researcher to start analysing the field material in parallel with the field work. That way the field work duration can be adjusted to fit with time periods that are meaningful from the point of view of the researched practices. 

\subsection{Space and Location}

Ethnographic field work should take place where the culture and practices that are studied take place. But where does software development take place? Software development is both anchored in the physical world, where software developers are located, and it takes place in the digital realm, where the software is located. Observations often need to combine physical observation and an understanding of how the physical actions relate to the development environment, digital documents, changes to and management of the source code, as well as the effects on the deployed software. 

Especially when cooperation and collaboration in distributed development is the focus of an ethnographic study, participatory observation on only one side risks neglecting the remote conditions for collaboration. To address this potential bias, multi-site ethnography emphasises the need to do ethnographic field work on all the locales involved. Multi-sited ethnography entails the researcher entering and being accepted at different sites which might require longer stays, e.g. one or more weeks, at all places where relevant project members or stakeholders are situated.

As an additional challenge, especially after the lockdowns due to the pandemic in 2020 and 2021, many companies moved from co-located or distributed development to hybrid work organisations, where developers partly work from home, leading not only to distributed but dispersed development, where every member of the team is in a different location \cite{DBLP:conf/xpu/SharpGM12}. Coordination and cooperation in these circumstances  more and more resemble Open Source development. Ethnographic research here has to follow the way software development is organised. 

One of the core challenges with researching at least partially virtual activities, is that it might be hard to follow an activity in a complex distributed development environment that both has physically visible parts, e.g. workstation layout and virtual elements, e.g. changes to code and deployed software. For example Begum \cite{Begum2020} describes the handling of a bug report: the bug report comes into existence through registration of the bug by a support engineer, followed by reproduction of the erroneous behaviour and a root cause analysis by a software engineer who prepares the discussion of the defect in a triage meeting, which in turn results in a strategy to fix the defect. The correction of the defect might again require collaboration between different teams responsible for different parts of the product. To understand what is discussed at the different meetings and to understand the rationale behind decisions might require at least a superficial understanding of the organisation of the source code.

Ethnographic field work here can follow an artefact, e.g. a bug report or the software architecture documentation, or it might focus on the activities of individuals, e.g. the software architect of (a part of) the software, or a group of developers. Different forms of ”virtual ethnography” have been developed as a way to observe internet based cultures (e.g. \cite{hine2008virtual, Kozinet2019, Pinketal2016} ).

Regardless of how well the field work is planned and prepared, even in smaller projects, the ethnographer will not be able to observe all relevant activities that are going on. Observations might need to be complemented with other data collection methods, like informal, in situ interviews, semi structured more formal interviews, or group interviews, e.g. in the form of a project retrospective \cite{DBLP:journals/infsof/DittrichL04}.

\subsection{Theoretical underpinning}

Ethnography often is seen as a bottom-up research method where the development of themes and concepts is based on reflective and inductive analysis. Theories are then developed by comparing results of several ethnographic studies. In order to relate ethnographic studies with each other, the themes and concepts developed in earlier studies have to be connected to the field work and analysis of the later studies, and this can be done via theory. One example is the social theory of learning by Lave and Wenger \cite{DBLP:books/cu/LW1991}: comparing findings from widely varying studies led to the formulation of a set of related concepts, a theory. This theory can then be used either as a focus for the field work in new ethnographic studies exploring social learning in different contexts, or to discuss the analysis of field material when peer learning emerges in the reflective and inductive analysis.

Theories can be seen as tools to think with rather than a claim for quantifiable correlations that are or can be supported by experiments. For example, the social theory of learning talks about ‘communities of practice’ \cite{DBLP:books/cu/LW1991} as an important facilitator to develop and maintain expertise in professional environments. ‘Communities of practice’ as a phenomenon can be identified both through the way members refer to each other and how knowledge is shared with new colleagues and among established members. Using the concept, for example, to understand the onboarding of novices in a software engineering team, can help to structure and focus the observations in the field. 

A danger of using theory from the outset, though, is that the focus that a theory provides can become a bias in the empirical work. It can prevent the researcher from paying attention to aspects of the observed practices that do not fit the theory. To prevent theory-based biases, the same techniques used for other biases can be applied (see this chapter, section 4.2).

Ethnography researches software development teams as sub-cultures or as social practices. Social theories that focus on social practices and how they reflexively constitute  social structure therefore might fit well with the observations and findings and help to develop relevant insights. There are several such practice theories. For example, activity theory \cite{engestrom1999activity} focuses on how human goal-directed activity is mediated by tools and also by rules, distribution of labour, and community. Activity theory has for example been used to better understand tool support for distributed development \cite{DBLP:conf/icgse/TellB12}. Dittrich et al \cite{DBLP:conf/sigsoft/DittrichMTLE20} and Dittrich \cite{DBLP:journals/infsof/Dittrich16}, use Schatzki’s \cite{Schatzki1996-SCHSPA-25} concept of social practices and Knorr Cetina’s \cite{cetina20005objectual} concept of epistemic practices as inspiration to understand the evolution and design of software development practices. Here, the way members of a software team adopt and adapt methods is at the core of the research. 

Distributed cognition \cite{hutchins1995cognition} emphasises the distributed nature of cognitive systems and focuses on collaborative work and the use of artefacts and representations. It results in an event-driven description which emphasises information and its propagation through the cognitive system under study. This has been used, for example, to understand how the structure of story cards and agile boards support collaboration \cite{DBLP:journals/ijmms/SharpR08} and how UX information is handled by agile development teams \cite{DBLP:journals/ijmms/ZainaSB21}.

The decision to consider theoretical frameworks can be taken based on the research intent, for example if the intent is to design and improve tools, the use of a theoretical framework that provides a vocabulary to identify and discuss the role of tools can be an advantage. As mentioned above, theory can also be used to interpret the data afterwards, e.g. Sharp et al. \cite{DBLP:journals/software/SharpRW00} used cognitive dimensions to analyse representations that were collected using distributed cognition as a framework. 

Giuffrida and Dittrich’s article \cite{DBLP:journals/infsof/GiuffridaD15} provides an example where ethnographic research is used to develop theory. They further develop concepts from coordination and communication theory to explain widely different success rates of student projects in distributed student projects. The initial analysis showed that successful projects used the Instant Messaging channel Skype provided to not only coordinate but also agree on coordination. In order to explain both what was going on and the differences, Genre theory and concepts based on articulation and coordination theory were combined and further developed. 

\begin{tips}
    {The Educators’ Corner: Tips for Teaching}
\begin{itemize}
    \item Spend time discussing what data to collect because observational data, e.g. meeting notes, is more subjective than other kinds of qualitative empirical data, e.g. transcripts of interviews or meetings.
    \item When discussing what data to collect, show the students that different levels of detail are possible and that they need to decide which level is relevant for them according to the research interest they have.
    \item When discussing what data to collect, show the students that the level of detail changes as the research progresses.
    \item The relationship with the gatekeeper and other informants needs ongoing management.
\end{itemize}

\end{tips}

\begin{question}{Exercise 5: Planning your ethnographic study}
\noindent Expand your study design developed so far with a concrete plan.
If you know the organisation and team you plan to collaborate with, describe the company, OSS community or other organisational anchor, the team, and the project. If you are working with a fictional ethnographic study design, describe the ideal organisation or OSS community, a team that would be typical for your research question, and an example of the kind of project you would like to engage with.
Then describe and motivate:
\begin{itemize}
    \item Whether you anticipate doing participatory or non-participatory observation.
    \item How long (elapsed time) and for how much time (percentage of a day or week) you anticipate the field work to go on.
    \item Where and how you expect the observation to take place.
\end{itemize}
\end{question}

\section{Implementing your ethnographic study}
\label{sec:4Implementation}

Planning can only prepare you so far for the research but here is where the ‘rubber hits the road’. As Fetterman \cite{fetterman2010ethnography} puts it: “The most important element of field work is being there - to observe, to ask seemingly stupid questions, and to write down what is seen and heard.”

\subsection{Gaining access and starting up}

After reading the previous sections, we expect the reader to have a clear understanding of what ethnography is, what kind of research questions it is best suited for, and the issues to consider when planning an ethnographic study. Furthermore, a “field site” to be studied has to be decided, i.e., a software development team, an open-source community, or organisation. However, as one can imagine, starting an ethnographic study is not only about “showing up” to observe what is going on at the field site. 

In general, it is important to have a key informant (also called a ‘gatekeeper’) that can facilitate the researcher’s access to the informants. This can be a project manager, team lead or a scrum master. This key informant is a person who is trusted, and possibly respected, by the informants and is willing to introduce the researcher to them so that they can start the data collection. In summary, this key informant negotiates access with the informants before the researcher goes to the field. Often, a good idea is to let this key informant introduce the researcher and suggest one or two initial members that are willing to be observed or interviewed. By interacting with these initial members, the researcher will have a chance to meet other team members and can negotiate with them additional opportunities for data collection. 

The key informant is also the person that the researcher will look for to ask clarification questions regarding the project being studied, its context, the diverse roles being played by the informants, the software development organisation, and the software being developed. This information often will also indicate documents and tools that the researcher should be aware of, and if necessary, the key informant will make the necessary arrangements so that the researcher can access these documents and/or tools. The key informant can be thought of as a facilitator who will allow the researcher to conduct the ethnographic study.

Part of starting up the actual field work is also to agree on the legal side of the collaboration and develop and sign cooperation agreements containing non-disclosure agreements and agreements about the intellectual property that might be developed as part of the research project. As part of starting up the field work, the researcher also needs to consider the research ethics relevant for this project and decide, maybe together with the key informant, on how to handle the informed consent. The latter topics are discussed further in section 7.
\begin{question}{Exercise 6: Observation}

\noindent Select a social context, which you can observe without disturbing the situation. This could be e.g. how people behave in the canteen when queuing for food and paying, or it could be how people behave on public transport.

Take notes on how people behave in this context. You might take different foci, e.g. how a specific person gets through the system, the role of artefacts like time tables or menus, events that change/disturb activity.

Formulate a text describing your observation to a traveller from outer space.

\end{question}

\begin{tips} {The Educators’ Corner: Tips for teaching}
\begin{itemize}
    \item Don’t impose your point of view. Let your student express their perspectives
    \item Challenge your student’s observations but gently
    \item Spend some time focusing on the use of language. How informants talk about their work or aspects of their organisation is very informative but attention to this kind of detail is not often emphasised in software engineering classes.
\end{itemize}

\end{tips}

\subsection{Handling your preconceptions}

One of the challenges of ethnographic research for a software engineer studying software engineering practices is that the researcher brings an education to the field that states how software should be developed. For example, having attended a software engineering course, the field worker might expect a ‘stand-up’ meeting to take place in situ by people actually standing up. However, in virtual software development, a stand-up meeting might mean that people sit in front of the screen communicating using a virtual meeting environment \cite{DBLP:conf/sigsoft/DittrichMTLE20}. Similarly, the textbook version of pair programming describes a “driver” and “navigator” role, but in practice pairing rarely adopts these roles formally, as the intensity of development results in a much more fluid exchange of activities \cite{DBLP:conf/xpu/PlonkaSSL11}. So how can a researcher handle such a situation, where the observed team is not doing things ‘by the book’? 

Ethnographic research talks about ‘bracketing’ prior assumptions and knowledge. In the above example it would mean accepting that pairing in practice changes shape when the focus is on developing code rather than following a pre-defined procedure, and that stand-up meetings may look different for different teams depending on their context. The research can then focus on how development is implemented in this context, what is the rationale for deviating from the textbook way of doing things, and how have these practices developed.

This concern of preconceptions extends to assumptions and definitions of terms. Even in the same team, different members of the team might use terms differently from each other and from the researcher, e.g. the term ‘prototype’ is used in different disciplines in widely varying ways. It is therefore important to pay attention to the language team members use when discussing development, and to challenge your preconceptions and assumptions on a regular basis to ensure that your understanding is correct. This allows the researcher to understand how and why methods and tools have been adapted and appropriated, but also the wording developers use can provide insights into their point of view regarding technical matters or development processes. For example how security incidents are referred to \cite{DBLP:journals/software/LopezTBLNS20}, or how development artefacts are talked about can indicate attitudes and concerns. The latter might also help the researcher to understand team structure and responsibilities, e.g. the tech lead might talk about his or her design, whereas the newby would talk about our design.

The importance of language is revisited below in Section \ref{sec:5Analysis} on analysis.

\subsection{During the study}

Observation is key to ethnographic studies and both participant and non-participant observation are legitimate forms of ethnography \cite{fetterman2020ethnography}. The data collected for observation is largely in the form of notes. Field notes are an important aspect of the work done by the ethnographer and the researcher must be prepared to take notes during the study. The question, though, is what notes to take? On the one hand, the researcher can document as many things as possible including the events, things people say, interruptions, event participants, documents being accessed or discussed, because at the start of the study it’s hard to know what will turn out to be significant. However, documenting everything will quickly become overwhelming. Textbooks sometimes recommend a template or schema for field notes. Such a schema can be very helpful and provide useful scaffolding at the start. Filling in the schema should not become the main purpose, though, and a tendency to just ‘tick the boxes’ needs to be resisted. The main purpose is, of course, to make sense of the team’s work practices and to document what is relevant to that aim. An alternative to schemas is to use a framework of questions. There are simple frameworks such as “who is there”, “where are they?” “what are they doing”, and more detailed ones such as the following based on Robson and McCarten \cite{RobsonMcCartan} (p. 328):

\begin{itemize}
    \item Space: What is the physical space like, and how is it laid out?
    \item Actors: What are the names and relevant details of the people involved?
    \item Activities: What are the actors doing, and why?
    \item Objects: What physical objects are present, such as story cards?
    \item Acts: What are specific individual actions?
    \item Events: Is what you observe part of a special event?
    \item Time: What is the sequence of events?
    \item Goals: What are the actors trying to accomplish?
    \item Feelings: What is the mood of the group and of individuals?
\end{itemize}

These are only intended to help keep a focus on relevant issues so that the ethnographer is not overwhelmed, and will evolve over time. 

Today, many researchers use computers for their field notes, which makes it easier to link to other field material such as (sample) documents, photos, and recordings. But handwritten notes are also popular and some find them a better way to engage with their surroundings. Each researcher will need to find their own preference.

However, as mentioned in sections 3 and 4, the focus of the study will evolve and the field notes will reflect that. For instance, after noticing that the software architecture influenced the interactions between the software teams, de Souza and Redmiles \cite{de2011awareness} started to pay more attention to the inter-team interactions because they were sources of tension between the different product teams. As the focus changes, ethnographers also adopt different data collection techniques, for instance, during the field work the researcher can decide to follow an artefact, e.g. a pull request, bug report, a presentation or the documentation of a piece of the software architecture. When this happens, the researcher should be prepared to ask for access to these artefacts which may mean a physical (print out) or digital copy of it. By following an artefact, the researcher might shine light on a different set of new events, activities and/or informants who often will bring new perspectives into the field site being studied.

Last but not least, it is important to note that field notes will contain both objective information such as quotes, timestamps of events, links to recordings and photos, and subjective information, such as the researcher’s reflections and interpretations of what is going on. Objective information is important, to anchor your emerging understanding of the work practices in actual observations. Verbatim quotations are extremely useful to present a credible report of the research. According to Fetterman \cite{fetterman2010ethnography}, “quotations allow the reader to judge the quality of the [ethnographic] work - how close the ethnographer is to the thoughts of natives in the field - and to assess whether the ethnographer used such data appropriately to support the conclusions.” 

The subjective information includes the researchers’ tentative interpretation of the events being observed, “hypotheses” about what is being observed, and questions the researcher asks themselves when collecting data. These reflections can evolve into important memos that in turn will develop into the parallel analysis of the data. It is, though, essential that the researcher clearly differentiates the subjective from the objective information in their field notes. This can be done by using different colours in a paper based field diary, or by splitting the page vertically or horizontally. Reflection and emerging hypothetical interpretations are, in qualitative research validated in the field. For instance, during data collection De Souza and Redmiles \cite{de2008empirical} started to wonder whether the phenomena observed would occur only with novice team members. To “validate” whether this is the case, they decided to observe and interview senior team members. Through this they found out that the phenomena was more general and occurred with team members of different experience. 

In some cases, the events being observed will unfold so fast that the researcher will not be able to take notes. In this case, ethnographers allocate time to record their memories in their field notes as soon as possible after the event, and by the end of the day at the latest. 

Nowadays, recording meetings is common and straightforward since some participants are online. In this case, it is important for the researcher to negotiate access to these recordings because they will be important data points for data analysis. They should be linked to the field diary so that the researcher later can connect them to the context in which they took place. However, being able to record meetings so easily is a potential disadvantage because it’s easy to collect a huge amount of data, and there is a tendency for the researcher to lose concentration if they know data is being captured through other means. 

In addition to observation, conducting ethnographic studies will likely employ a range of data collection methods including interviews and document analysis. Interviews are a chance to clarify meanings and the history of what has been observed. Such interviews might be formal or informal. Informal, or impromptu, interviews are often in situ and opportunistic; they will often occur in the context of an observation. In other words, the ethnographer, after witnessing an event that seems interesting, surprising, or even expected, will use the opportunity to engage in a quick conversation with the informant to find out more details. These informal interviews will occur when opportunities arise, for instance, during a break or while waiting for a meeting. All this is done having in mind that the role of an ethnographer is to understand the member’s point of view. Formal interviews require an agenda and a negotiated date and time, but may be the only opportunity to speak with some informants or to explore some kinds of issue. 

In our experience, a good ethnographer strikes a balance between conducting formal and informal interviews. What this balance looks like depends on the context and the flexibility of the informants’ working day. Too many informal interviews with the same informants can become irritating. Similarly, conducting formal interviews means asking people to allocate time for the conversation. However, checking information and making sure that observations and reflections are well-founded are fundamental.

A good ethnographer also needs to pace themselves, plan their days carefully and be systematic about keeping records. This is particularly significant because data collection will go on for several days or even weeks, and it is easy for memories to fade or become confused. The researcher needs to be constantly reflecting about what data is being collected, including around events, conversations, and artefacts, and revisit the emerging analysis concepts in light of the new observations. Ethnographers need to remember to avoid judging the work being observed and to keep focus on the informants’ point of view, i.e., why they do the things they do. Being careful about data collection also means collecting data from different informants to gather different perspectives while, at the same time, not being a nuisance.

Finally, it is important to emphasise that ethnographic research, similar to other qualitative research methods, requires cycles of data collection and analysis. In other words, ethnographic research requires negotiating continuous and evolving access to the site. As discussed in the next section, during the data analysis new foci might emerge and require new directions for data collection, such as access to new informants or new events or new documents. So the ethnographer must be continuously negotiating access to these during the research process.

\subsection{Going Native}

A key ability for an ethnographer is to be able to view activities as “strange”, so that they can question what is happening and uncover the significance of what is observed. But it’s also key to be able to get inside a culture and elucidate the members’ perspective. Keeping a balance between strangeness and familiarity can be challenging. If the balance tips too far in the direction of familiarity and it becomes hard to view things as strange then the researcher is said to have “gone native”. 

Here regular debriefing meetings with supervisors and colleagues are important to keep the analytic distance. 

\begin{warning}{The Educators’ Corner: Pitfalls}

\begin{itemize}
    \item Being judgmental about observed practice SO focus on why it is the way it is.
    \item “Going native” meaning that the way informants behave and achieve things become ‘normal’ SO talk to other researchers and hold debriefings to help you regain the questioning mindset.
    \item Losing your “strangeness” perspective and not knowing what needs to be told for others to understand your interpretations SO try explaining what is happening to an alien.
    \item Human tendency is to gravitate towards people you like but in the field it’s important to attend to all informants SO be aware of your own preferences and biases.
    \item It is easy to focus on a small number of informants because you become comfortable with them SO make sure you choose a range of informants to get different perspectives, i.e., novices and experts, male and female, informants with different roles, etc.
    \item Observation can result in lots of data, some of which is not relevant SO focus on what to pay attention to and choose carefully what data to collect.
    \item Assumptions or pre-conceived ideas can lead to misunderstandings SO challenge your own assumptions and ask dumb questions to establish that your understanding is correct.
    \item Recording meetings and interviews is straightforward and easy but can result in huge amounts of data and loss of concentration SO keep taking field notes even when the interactions are being recorded.
\end{itemize}

\end{warning}

\subsection{Leaving the field}

Eventually, the researcher will have to leave the host organisation or the open source community, because the software development project finishes, the research project ends, or because the researcher has learned what could be learned from the ethnographic field work. Leaving the field should be done in a way that allows the researcher to come back to ask clarifying questions and open up for future collaboration on new mutually interesting research questions. It is therefore important that the researcher informs all of the informants that they are leaving the site and what will happen next with their data and its analysis. 

Saying farewell to the team can be combined with a presentation of (preliminary) findings and insights that can also be used as an occasion for member checking. Member checking is an important aspect of any qualitative research (see also Section 5). It involves presenting the results of the research to the informants so that they can “validate” them. In ethnographic research, one way to do this is to formally present the results to the key informants who facilitated the ethnographer’s access (see Section 4.1), or even to the whole development team. Another way is through informal presentations during the final days of data collection, such as during a break, while waiting for a meeting, etc. The format in which this is done varies depending on the team, project, and overall context. On the one hand, feedback from the participants is important for the trustworthiness of the findings. On the other hand, presenting the findings can also serve as a ‘thank you’ for the team and can support the reflection and improvement of their own practices. If an organisation or team is interested enough in having a researcher on site then they will also be interested to know what insights have been found. As one of our collaborators once said “the problem with running a project is that you’re running, but having you here allowed us to stop and reflect”. 

\begin{question}
    {Exercise 7: Prepare for the first day for your field work}
\begin{enumerate}

    \item Identify your key informant: \\Add to your study design a section motivating this and describing how the contact can be established.
    \item  Design your field diary: \\Decide whether to use an electronic diary or a paper diary. Decide on how to separate observations and reflections. Decide whether to use a schema for your field diary and if so, which one.
    \item Add a few paragraphs to your study design motivating and presenting your choice.
\end{enumerate}

\end{question}

\begin{warning}{The Educators’ Corner: Pitfalls}

\begin{itemize}
    \item Believing that you’re in the field to improve practice SO focus on understanding why it is the way it is.
    \item Having preconceived ideas such as for a new tool or process improvement SO bracket your assumptions (see Section 4.2).
    \item Getting observations mixed up with your interpretations SO structure your diary to make them distinct.
    \item Making too much of the detail of everyday practice rather than why things are the way they are SO get into the habit of regularly asking yourself “why”.
\end{itemize}

\end{warning}

\section{From Data to Text}
\label{sec:5Analysis}

The first part of the term Ethnography (“ethnos”) means culture, while the second part (“graphy”) refers to the writing, the presentation of information about the culture or practices observed. When using ethnography as a research method in software engineering, the presentation of the results is in the form of articles, and less often in the form of a monograph. The challenge often is to reduce the rich experiences from the field into what can be presented in the format of a conference or journal submission. The more organised the ethnographer, the easier the task of analysing the collected data \cite{fetterman2020ethnography}. We first present the analysis process and thereafter discuss how to relate it to the software engineering discourse.

\subsection{Reflective and inductive analysis}

The analysis of ethnographic data is a reflective and inductive iterative process that starts in parallel to field work. It is possible and even recommended that ethnographers write temporary notes (often called memos) about what is observed in the field. The information captured in memos will be used backwards to identify where else a similar pattern showed up in the field work so far, and forwards to inform the following period of data collection. For instance, as mentioned in the previous section, de Souza and Redmiles \cite{de2008empirical} interviewed experienced developers to find out whether their experience was similar to novice developers. 

These memos are also an important starting point for a more formal and systematic analysis. But where to start the analysis is a conundrum when faced with so much rich data. One useful technique is to look for “rich points” \cite{Agar1996}. These are surprises or notable insights that researchers experience while in the field. Often they may manifest as initial impressions when entering a new event or context (although any such impressions should be systematically validated). These may lead to narrative accounts of everyday practices, or to identifying patterns that appear across situations. The former is not simply a description of what was observed, but includes a synthesis and interpretation of the data. Being careful with language both during analysis and in the ethnographic account is very helpful here: “certain birds can solve problems better than some dogs” carries a different meaning to “birds are smarter than dogs”. In the process of capturing and reporting findings, it can be easy to summarise or abstract something that accidentally leaves out important context or nuance.

Ethnography, like other qualitative research methods, often employs coding techniques to identify patterns across situations, i.e., data are micro-analyzed (line-by-line) to identify codes. When coding, an ethnographer is actually making sense of all the data collected. These codes are put together under a more abstract, high-order concept to explain what is going on \cite{clarke2021thematic}\cite{Hoda2024grounded}. For instance, \cite{de2008empirical} de Souza and Redmiles \cite{de2008empirical} created a code to indicate the concept of “back merges”: the fact that a software developer had to incorporate code changes from other developers into their own workspace before integrating their code into the main codebase. Codes are created to minimise the number of elements that the researcher needs to consider. After creating codes, and often in parallel to them, codes can be grouped together to suggest a category, an even high-order concept. In this case, de Souza and Redmiles \cite{de2008empirical} created the category backward impact management to indicate how a software developer continuously assesses the impact of the work performed by other developers on their own work, and the appropriate actions developers adopted to avoid such impact. In other words, a “back merge” (code) is a form of backward impact management (category). 

However, coding is not necessarily only informed by the data. For example inductive codes might be combined with codes that relate observations to theoretical concepts. Likewise, there is no requirement for code coverage, as the data collection especially in the introduction of the observation is not yet focussed on what emerges to be the main theme. For a more detailed coverage of qualitative analysis see the chapter by Treude in this volume and the Further Readings provided in section 7.

In general, the goal of the analysis is to invoke all data collected (field notes, documents, code fragments, informal and formal interviews, etc) to draw a clear picture of how the field site works, i.e., why the software developers do what they do. When envisioning this big picture the ethnographer’s attention will often wander through the data and will make logical leaps that lead to interesting insights. However, they must also “backtrack to see whether the data will support these new ideas or invalidate them” \cite{fetterman2010ethnography}. Ethnographers and qualitative social scientists talk about a ‘dialogue with the data’. During analysis the researcher moves between formulating reflections and interpretations leading to codes, which then are tested with the empirical data \cite{Pawluchetal2017}. This also includes data triangulation, i.e., exploring different sources of data and making sure they support or at least don’t refute the result. For instance, the results from observation can be combined with the analysis of interviews, or  documents. In addition, an experienced ethnographer will use “member checking”, i.e., will share their writing, memos, or any form of temporary results with trusted informants to gather feedback and find out whether their understanding of what is going on in the field site is recognised by the informants. It is important that the description of what was said and seen in the site be recognized by the informants, even if they do not fully agree with the interpretation of these results \cite{fetterman2010ethnography}. 

\begin{tips}{The Educators’ Corner: Analysing data together}

\noindent One way data analysis is taught in social sciences is to analyse data together. Here researchers take an anonymised version of their own field material, an interview transcript, an instant messaging protocol, or a video or audio recording and analyse it with others.

Of course it would not be possible to reconstruct the full ethnographic experience. However, in joint analysis, the teacher could point to relevant cues that connect the specific part of the field material to the broader context of the empirical work and that way illustrate how the brought picture gained through ethnographic field work allows them to then make sense of the specific document or conversation.

\end{tips}

\subsection{Writing Ethnography for Software Engineering Audiences - Reporting the Results}

The results of analysis need to be related to software engineering literature to argue how the findings lead to new insights for the software engineering community. This has traditionally been difficult to achieve, as many software engineering researchers saw it as the goal of software engineering to improve industrial software development practice rather than to understand why practitioners do things the way they do. In recent years, though, this has started to change. Ethnography and qualitative empirical research are more accepted as a means to understand development practices better, and hence to improve understanding which can then inform the development of new tools and techniques.

Although the overall theme of the research is likely to have been established before field work starts (if only to inform the choice of field site), the specific interesting aspects that support new insights for the software engineering community will evolve during field work and analysis. Because of this, the researcher might have to complement the usual initial literature study with a short field study to establish the value of their chosen research focus, especially for a Master’s or PhD project. 

Sometimes, additional analysis or even additional research is necessary to estabish an argument for the relevance of the results. In this case, ethnography is used in combination with other studies using different techniques. An example of this is described in Plonka et al. \cite{DBLP:journals/ijmms/PlonkaSLD15}. In this research ethnographic field work was complemented by video recordings of pair programming sessions. The analysis of ethnographic data led to the identification of a number of ways in which knowledge transfer takes place during pair programming. This in turn was supported by an in depth interaction analysis of the video recordings, which both illustrated and deepened the understanding of knowledge transfer in pair programming and led to some practical recommendations \cite{DBLP:phd/ethos/Plonka12}.

An example where additional research developed into an empirical (sub)project in its own right is Unphon and Dittrich’s article \cite{DBLP:journals/jss/UnphonD10}. They present an interview study that was based on a finding in the above-mentioned ethnographic-grounded action research study. Unphon termed the observation of the technical lead communicating the architecture through one on one discussions ‘walking architecture’. The interview study established that this practice was wide-spread in software product development organisations and also uncovered some of the rationale behind these practices.

When writing up the study, the method section needs to state the basic information about the site and duration of the study. But also a detailed account is needed of the field work methods, the analysis process and the measures implemented to assure the trustworthiness of the qualitative research. Often reviewers will ask for numbers, e.g. hours/days of observation, number of documents analysed, and even the size of the system being developed by the team. These can be included to characterise the study itself, but do not be tempted to analyse these numbers quantitatively as part of the results. Also the measures implemented to assure trustworthiness (see subsection above \cite{RobsonMcCartan}) need to be presented in detail and examples of how these methods were applied in this study will often be expected.

One dilemma is the conflict between the need to provide rich descriptions and the result orientation of the software engineering community. Passos et al. \cite{DBLP:conf/esem/PassosCDM12} also mention this as one of the challenges of ethnography in the context of software engineering. Rich descriptions allow the reader to follow the inductive reasoning that led to the findings that support the insights. Software engineering reviewers often find these rich descriptions too long and uninteresting. Here the qualitative researcher will have to strike a balance between the quality of the account and the requirements of the community. Using tables, charts or other techniques to summarise context and findings can help establish such a balance. 
\begin{tips} {The Educators’ Corner: Tips for teaching}

\begin{itemize}
    \item Writing as a part of interpretation and reporting is not always comfortable for those with an SE background, so offer writing guidance.
    \item Language is important in the data so include some analysis exercises that focus on the use of language.
\end{itemize}

\end{tips}

\begin{warning}
    {The Educators’ Corner: Pitfalls}

\begin{itemize}
    \item First impressions can overshadow later observations SO look for confirming AND disconfirming evidence.
    \item Succumbing to confirmation bias SO seek confirming AND disconfirming evidence for your interpretations.
    \item Getting hung up on too much detail SO sit back and consider the wider picture.
    \item Making too much of the detail of everyday practice rather than why things are the way they are SO get into the habit of regularly asking yourself “why”.
\end{itemize}

\end{warning}

\section{Ethnography and Research Ethics}
\label{sec:6Ethics}

As part of ethnographic field work, the researcher tries to become part of the real world environment that he or she researches. This means that the researcher will get to know about aspects of the team members’ work and life that potentially can lead to harm when exposed in the wrong contexts. Reflection on research ethics therefore needs to be an integral part of ethnographic research.

In many universities empirical research designs need to be approved by an ethics board. These boards might have diverging requirements from the considerations below. In some aspects they might ask for more measures for assuring ethical implementation of the research design, in some aspects the considerations below might appear more rigorous. Our discussion addresses what we think is needed to act ethically and responsibly as an ethnographic researcher in an industrial software engineering environment.

As empirical software engineering researchers, we often cooperate with project teams or IT departments that are part of a company. As both the cooperating team and the management can be expected to read our articles, the results of the ethnography in itself might lead to problems for the team or department in the organisation. Lucy Suchman argued in her seminal article ‘Making Work Visible’ \cite{DBLP:journals/cacm/Suchman95a} that the presentation of how work actually takes place in an industrial environment can, on its own, be a highly political act. However, the company itself can also face negative effects of the research.

In ethnographic research, ethics is not only about the effects of the research when published. How the researcher acts in the field and how their loyalties are perceived, will affect the trust of the software team or project they are working with. How ethical issues are handled needs to be part of the planning of a research project.

The researcher needs to reflect on the impact of the research on different levels:

\begin{itemize}
    \item The individual team member might face repercussions if it becomes visible that they do not work in adherence to the project’s or the company’s methods or policies. 
    \item The software team could be affected by having their internal work practices exposed. 
    \item The relationship between team and management might suffer, if the team perceives the researcher as a control agent reporting to management.
    \item The company could suffer from exposure of internal work practices, e.g. with respect to the reputation and, through that, trust of customers and investors. Further, the exposure of the plans and development of innovative services might diminish the competitive advantage the company tried to achieve.
\end{itemize}

Although it is often possible to anonymise the company and the project to the wider audience and abstract from the specific software developed, the anonymisation of the identity of the team and of team members with respect to the company organisation might not be possible. 

Handling your relationship with the individual developers, the team, and the company in a responsible manner therefore needs to be addressed in a number of ways.

The researchers, one way or another, need to acquire informed consent. This means that team members are aware of the researcher and the research interest. They are aware of how the data collected will be used, who has access to it prior to publication and how the use of direct citations or observations will be communicated to them before disclosure inside the company or publication of research results. Informed consent might be achieved by asking the team members to sign a document. It also can be achieved by the researcher presenting themselves and detailing these topics, for example at the beginning of a team meeting or through an e-mail. 

Prior to publication and prior to the management clearing publication from a NDA point of view, citations of team members and the presentation of their behaviour need to be member-checked with the individual team members. This has two purposes: to make sure that the researcher understood situations and behaviour in an adequate way; and to allow the individual team members to control what is disclosed about their behaviour to management. To this end, it is common to create excerpts from the analysis section of publications or theses and share them with the relevant team members. 

Open negotiation and communication about how findings are revealed to both the management and the academic public, is crucial for trust between team members and researchers. In many cases, the company’s management itself is not interested in being perceived as using researchers to ‘spy’ on their employees. In one of our projects, the management of the company took up the issue by asking in the beginning of the collaboration, whether we researchers plan to act as a modern version of the ‘time and motion’ recorder. In a Tayloristic approach to the organisation of industrial production, researchers recorded the time it took for the individual operations of an assembly. The ‘time and motion recorder’ since then has become a symbol for control in many European countries. The discussion that followed this comment from our collaborators helped us to clarify the role of the researchers. Further, it is important that the management is aware of the commitment of the researchers to not disclose findings if the employees and the team disagree.

In some cases, especially when combining ethnographic research with action and design research, the researchers might consider co-authoring research results together with (some of) the involved practitioners. This requires the contents, conclusion, and structure of the publication to be discussed with the relevant practitioners. During the writing process, the researchers should then as much as possible involve the practitioners, so that all authors agree on the text, especially the findings and conclusions.

Non-Disclosure Agreements and contracts handling Intellectual Property detail the procedures to make sure the interests of all parties are considered before results are published. Most collaboration agreements between companies and universities include procedures for review and acceptance of publications containing the research results. However, we recommend taking up this topic with the company contact early and discussing any aspects of the project that can become critical for the company if disclosed. Some companies might prefer to not be mentioned by name in the article, although a sufficient description of the research context may result in identification anyway, e.g. describing the collaborator as “a large multi-national telecomms company based in Finland” suggests only a few possible companies. Sometimes certain aspects of the software under development and the project could provide information competitors could use to harm the company. In most cases, the researchers can find ways to disguise critical information.

Ethnographic research is sometimes used as a first step in action or design research. In such cases, we recommend extending the established practices to handle relationships with the developers, the team and the company to the other research activity. Dittrich et al. \cite{DBLP:journals/ese/DittrichREHL08} propose designing their participatory action research approach in a way to keep the members’ perspective throughout the action research cycle.

\begin{question}{Exercise 8: Consider research ethics}

\noindent Review your ethnographic study design from a research ethics point of view and revise it based on the issues discussed in this section.

\end{question}

\section{Final comments and Further reading}
\label{sec:7}

At the start of this chapter we provided a simple answer as to why ethnography is different to other qualitative methods. Having read through the chapter and hopefully completed the exercises along the way, we hope that you have a clearer and more holistic answer to this question. Ethnography is characterised by focusing on the members’ point of view and understanding why things are the way they are. Understanding current practice helps to inform decisions about future practice, processes and tools. But it’s not the specific practice that is relevant; it’s the significance of that practice. Ethnography is good at answering “how” and “why” questions and makes certain demands of the researcher, such as focusing on everyday detail and writing thick descriptions. It has been used to address research questions in four contexts: investigations of social and human aspects, to inform tool design, to inform process change and to complement other research methods. So we invite you to consider this question again, this time for yourselves:

\begin{question}{Exercise 9: Ethnography and other research methods}

\begin{enumerate}
    \item 
What are the key differences between ethnography and:

\begin{enumerate}
    \item Grounded theory
    \item Action research
    \item Case study?
\end{enumerate}

\item How might an ethnographic study complement a research programme that involves:

\begin{enumerate}
    \item Experiments
    \item Survey
    \item Systematic literature review
\end{enumerate}
\end{enumerate}
\end{question}

\noindent The following are suggestions for further reading on this topic:
\\

\noindent \textbf{\footnotesize Fetterman, D.M.: Ethnography: Step-by-step. Sage publications (2020)}  

This is a very practical book about ethnography and although written with a social science audience in mind, it is accessible to readers from a wide range of backgrounds.
\\

\noindent \textbf{\footnotesize Hammersley, M., Atkinson, P.: Ethnography: Principles in practice. Routledge (2019)} 

This article is written from a social sciences perspective, and aims to take an objective view of ethnography and its limitations. It deepens and problematizes some of the issues raised in this chapter and introduces some others. 
\\

\noindent \textbf{\footnotesize Pink, S., Horst, H., Postill, J., Hjorth, L., Lewis, T., Tacchi, J.: Digital Ethnography: Principles
and Practice. SAGE Publications Ltd, United Kingdom (2016)} 

This text provides an introduction to digital ethnography and how different media and technologies affect the ethnographic research endeavour, both through their effects on the world being studied and the opportunities that technology provides for the researcher.
\\

\noindent \textbf{\footnotesize Randall, D.W., Harper, R.H.R., Rouncefield, M.: field work for Design - Theory and Practice.
Computer Supported Cooperative Work. Springer (2007) }

The book presents a social science perspective on ethnography in the context of Computer Supported Cooperative work. It explains ethnography and how it can be brought to bear on design of software to support cooperation. Though this is different from using ethnography in software engineering, it supports the possibility to use ethnography in different disciplinary contexts and reflects on the challenges of doing so.
\\

\noindent \textbf{\footnotesize Zhang, H., Huang, X., Zhou, X., Huang, H., Babar, M.A.: Ethnographic research in software
engineering: a critical review and checklist. In: 
Proceedings ESEC/FSE 2019, Tallinn,
Estonia, August 26-30, 2019, pp. 659–670. ACM (2019). } 
As research is progressing, new ethnographic studies are published continuously. This is one recent literature review: 
\\

\noindent \textbf{\footnotesize Sharp, H., Dittrich, Y., de Souza, C.R.: The role of ethnographic studies in empirical software
engineering. IEEE Transactions on Software Engineering 42(8), 786–804 (2016)} 

This article is a companion to this chapter. It introduces the role of ethnography in empirical software engineering through a set of ethnographic studies that have been conducted with software engineering as their focus.

\begin{acknowledgement}
Thanks to our colleagues and students who have joined us in our daring endeavour into researching everyday practices of software engineering. Together with you we learned what was challenging in ethnographic studies.
Thanks to the software engineers who collaborated with us. You have taught us so much about how software engineering actually takes place. 
Thanks also to the reviewers and editors of this volume for your constructive feedback and support.
\end{acknowledgement}

\bibliographystyle{spmpsci}
\bibliography{EthnographyChapter}

\end{document}